\documentclass[nofootinbib,showpacs,preprintnumbers,amsmath,amssymb,floatfix]
{revtex4}
\usepackage{epsf}

\begin{document}


\title{Comment on double-logarithmic corrections to decays of electroweak bosons}

\vspace*{0.3 cm}

\author{B.I.~Ermolaev}
\affiliation{Ioffe Physico-Technical Institute, 194021
 St.Petersburg, Russia}

\begin{abstract}
We present a short comment on resummation of double-logarithmic (DL) contributions to decays of
$W$ and $Z$ -bosons into charged fermion pairs. We show that resummation of DL corrections
to the $Z$ -boson decay leads to
standard Sudakov form factor with the negative exponent independently of the boson is on-shell or
off-shell. In contrast, the Sudakov exponent for the $W$ - decay is negative (positive) when
the $W$-boson is on-shell (off-shell). In this connection we consider applicability of the Dispersion
Relations to the case when the imaginary parts contain infrared-divergent contributions.

\end{abstract}


\maketitle

\section{Introduction}
The vertex $\Gamma_{\lambda}$ of the electromagnetic interaction of fermions includes two form factors,
$f(q^2)$ and $g(q^2)$:
\begin{equation}\label{vertex}
\Gamma_{\lambda} = \bar{u} (p_2) \left[ \gamma_{\lambda} f(q^2) -
\frac{\sigma_{\lambda\nu} q_{\nu}}{2 m}g(q^2)\right] u(p_1) ~,
\end{equation}
with $q= p_2 - p_1$. Form-factor $f(q^2)$ at large $q^2$ was
studied first in  Ref.~\cite{sud} where the double-logarithmic (DL)
contributions were discovered in the QED context. Contrary to
other radiative corrections, DL contributions depend on the infrared cut-off.
Results of resuming such terms to all orders in the
coupling(s) are written in an exponential form, with a negative
exponents, so that the exponentials fall
when the total energy grows. This is known as the Sudakov suppression. It means the
steep suppression of non-radiative processes compared to
the processes where bremsstrahlung is allowed.
After that the Sudakov exponentiation for the form factor $f(q^2)$ was obtained
in  the QCD context\cite{quarkff} and in
Standard Model\cite{flmm}. Resummation
of double-logarithmic QED and QCD corrections to the form factor $g(q^2)$ was done in Ref.~\cite{et}.
Asymptotics of this form factor has also been considered in the
recent paper \cite{ciaf}.

In the present paper we consider resummations of DL corrections to decays of the
electroweak bosons into fermions: $Z \to f \bar{f}$ in Sec.~2 and $W \to f' \bar{f}$ in Sect.~3. Besides giving
explicit results for such resummations, we find interesting to consider in more detail the following technical aspect:
quite often it is convenient to calculate amplitudes of various processes, calculating first their imaginary parts and then
using the Dispersion Relations. In this approach one should first calculate imaginary parts
(discontinuities) of the involved Feynman graphs and then use the Dispersion Relations. For the processes we
consider below we use the Dispersion Relations with one subtraction and with logarithmic accuracy:
\begin{equation}\label{dr}
M(s) = \frac{s}{\pi} \int_{s_0}^{\infty} d s' \frac{\Im M (s'/s_0)}{s' (s' - s)}  \approx -
\frac{1}{\pi} \int_{s_0}^s d s' \frac{\Im M (s'/s_0)}{s'}~.
\end{equation}

The imaginary parts are free of the ultraviolet divergences, so the use of the
Dispersion Relations makes possible to avoid the procedure of
renormalization and therefore considerably simplifies the calculations.
The second very attractive feature of this approach is that the graphs with zero imaginary parts
can be left out from analysis since very beginning. In particular, this point was observed and used by L.N.~Lipatov
in Ref.~\cite{ln} and in subsequent papers. It proved to be especially effective for analysis of many-particle
processes in QED and QCD (see e.g. Refs.~\cite{kfl, efl, el}). However, such a selection of graphs
should be done very carefully otherwise it can lead to wrong results.
We give an example of  such situation in Sect.~3 when calculate the $W$ -decay.

\section{DL corrections to the decay $Z \to f \bar{f}$}
In this section we consider the decay of $Z$ -boson into the fermion-antifermion pair $f \bar{f}$, with the both
$Z$ -boson and produced particles being on-shell and assuming that their mass $m$ is small compared to the mass
$M$ of the $Z$ -boon.

\subsection{First-loop DL contribution to $Z \to f \bar{f}$}

Through the paper we use the Feynman gauge. So, the only first-loop Feynman graph contributing to the
decays $Z \to f \bar{f}$  in DLA is depicted in Fig.~1. 
We assume that the produced fermions have non-zero
electric charges and therefore the dashed line in Fig.~1 corresponds to the
virtual photon. Obviously, neither $Z$ nor $W$ -exchanges can yield DL contributions. The
standard way for integration over
the virtual photon momenta $k$ involves using the Sudakov parametrization\cite{sud}. In general, it means representation
of $k$ as
\begin{equation}\label{sud}
k = \alpha p'_1 + \beta p'_2 + k_{\perp}.
\end{equation}

The massless momenta $p'_{1,2}$ in Eq.~(\ref{sud}) are
$p'_1 = p_1 - x_1 p_2,~~p'_1 = p_2 - x_2 p_1$, with  $x_1 = p_1^2/(2p_1p_2),~~x_2 = p^2_2/(2p_1p_2)$.
When the produced fermions are on-shell, $p^2_{1,2} = m^2$.  Through the
paper we will keep the notation $w \equiv 2p_1 p_2 = M^2 - p^2_1 - p^2_2 = M^2-2m^2 \approx M^2$.
DL contribution of the graph in Fig.~1 arrives from the region of soft $k^2$, so the infrared cut-off
$\mu$ has to be introduced. After that the first-loop expression $M_Z^{(1)}$ for the decay is
\begin{equation}\label{zint}
M_Z^{(1)} = M_Z^{Born} \frac{\alpha Q_f \bar{Q}_f}{4 \pi} \int \frac{d \alpha d \beta d k^2_{\perp}}
{-2 \imath \pi} \frac{2 w}{(w\alpha + x_2 w \beta + k^2+ \imath \epsilon))(w \beta + x_1 w \alpha
+ k^2 + \imath \epsilon))(k^2 + \imath \epsilon)}~,
\end{equation}
with $k^2 = w \alpha \beta - k^2_{\perp}$. In Eq.~(\ref{zint}) we have
used the following notations: $M_Z^{Born}$ is the same amplitude in the Born approximation,
$Q_f$ and $\bar{Q}_f$ are the
factional charges of the fermion and antifermion respectively ($\bar{Q}_f = - Q_f$).

Integrations in Eq.~(\ref{zint})  can be performed with the standard means
and the result is given by the following expression:

\begin{equation}\label{zloop}
M_Z^{(1)} =  M_Z^{Born} \frac{\alpha Q_f \bar{Q}_f}{4 \pi} L .
\end{equation}
In general,
the double-logarithmic factor $L$ is  (for detail see e.g. Ref.~\cite{egot})
\begin{equation}\label{l}
L =  \ln^2 (M^2/\mu^2) - \Theta (m^2_1 - \mu^2) \frac{1}{2} \ln^2(m^2_1/\mu^2)
- \Theta (m^2_2 - \mu^2) \frac{1}{2} \ln^2(m^2_2/\mu^2)
\end{equation}
where for the $Z$-decay we should put $m_1= m_2 = m$.

\subsection{Total resummation of DL contribution to $Z \to f \bar{f}$}

DL contributions from higher loops arrive from the Feynman graphs similar to the one in Fig.~1 however
with more photon propagators connecting the fermions lines. Both ladder and non-ladder graphs are essential.
According to , The total resummation of their DL contributions can be done
either with direct analysis of the involved Feynman graphs like in Ref.~\cite{sud} or
with composing and solving an appropriate evolution equation (see e.g. Ref.~\cite{efl} or the
overview \cite{egt}).
It leads to exponentiation of
the first-loop expression, so the result is
well-anticipated and can be reproduced from many sources:
\begin{equation}\label{mz}
M_Z = M_Z^{Born} \exp \left[ \frac{\alpha Q_f \bar{Q}_f}{4 \pi} L\right] =
M_Z^{Born} \exp \left[- \frac{\alpha Q_f^2}{4 \pi} L\right] ~.
\end{equation}

\section{DL corrections to the decay $W \to f' \bar{f}$}

Calculating DL contributions to decays of the charged bosons is less trivial from the technical point of view,
even in the first loop.
Surely, the graph in Fig.~2 has to be accounted in the first place. The dealing with it is quite similar to
the contents of the previous Sect and leads to the following result:
\begin{equation}\label{wloop}
M^{(2)}_W = M_W^{Born} \frac{\alpha Q'_f \bar{Q}_f}{4 \pi} L~,
\end{equation}
with $L$ given by Eq.~(\ref{l}) and $Q'_f$ ($\bar{Q}_f$) being the electric charge of the fermion $f'$ (antifermion
$\bar{f}$. Let us remind that for the $W$ -decays $m_1 \neq m_2$ in Eq.~(\ref{l}) because $f \neq f'$.
The fundamental difference between Eqs.~(\ref{wloop}) and (\ref{zloop}) is that the
products of the fractional charges in them have the opposite signs. Indeed, the
 factor $Q_f \bar{Q}_f = - Q^2_f < 0$ regardless of the sign of $Q_f$
 while $Q'_f$  and $\bar{Q}_f$ in Eq.~(\ref{wloop}) have the same sign
 (positive or negative, depending on the $W$-boson sign)
 and therefore $Q'_f \bar{Q}_f$ is always positive. In contrast to the $Z$ -decay, there is one more contribution
 to the $W$ -decay in the first loop. It comes from the graphs in Fig.~3. In principle, there could be an argument claiming
 that this graph cannot yield a DL contribution when the initial $Z$ -boson is on-shell:
 let us calculate any of these graphs, using the Dispersion Relations
 approach, then in the first place we should calculate its its imaginary part (with respect to $w$), $\Im M^{(3)}_W$.
In order to calculate it, one should cut the both graphs in Fig.~3 i.e. replace the propagators
$\left[(p_1 + p_2 -k)^2 - M^2 + \imath \epsilon\right]^{-1}
(k^2 +  \imath \epsilon)^{-1}$ of the
cut  $Z$ -boson and photon  by $(- 2 \pi \imath)^2 \delta\left[(p_1 + p_2 -k)^2 - M^2 + \imath \epsilon\right]
\delta (k^2 + + \imath \epsilon)$. It puts the $Z$ -boson and photon on the mass shell.
Eq.~(\ref{dr}) demonstrates that the single-logarithmic (SL) terms in $\Im M^{(3)}_W$, being substituted into the Dispersion
Relations, yield DL contributions to $M^{(3)}_W$.
On the other hand, there is the well-known knowledge
following from the energy-momentum conservation:
any on-shell particle cannot
emit or absorb an on-shell photon. It means that the
imaginary parts of both graphs in Fig.3 are zeros when the external $Z$ -boson is on-shell and
therefore SL contributions cannot come from these graphs.
However, it is easy to see that this argument fails for the SL contributions:
Indeed, SL contributions to the imaginary parts of these graphs
contain the infra-red (IR) divergency
when $k^2 = 0$, i.e. when the cut photon is on-shell. In order to regulate it, an infra-red
cut-off $\mu$ must be used which is equivalent to providing the photon with a mass (virtuality).
Emitting or absorbing a virtual photon by an on-shell particle is allowed, so graphs in Fig.~3
are not forbidden to contribute DL terms. The other, non-SL contributions are infrared stable, so when they are considered,
the cut photons in Fig.~3 can be really on-shell and therefore non-SL contributions to
$\Im M^{(3)}_W$ are absent when the external $Z$ -boson is on-shell.
The DL contribution of the graphs in Fig.~3 is
 \begin{equation}\label{waddloop}
M^{(3)}_W = - M_W^{Born} \frac{\alpha }{8 \pi} [Q_W Q'_f L_a + Q_W \bar{Q}_f L_b]
 \end{equation}
where $Q_W$ is the $W$ - fractional charge, $Q_W = Q'_f + \bar{Q}_f = \pm 1$, and
\begin{equation}\label{lab}
L_a = \ln^2 (M^2/\mu^2) - \Theta (m^2_1 - \mu^2) \ln^2 (m^2_1/\mu^2),~~~
L_b = \ln^2 (M^2/\mu^2) - \Theta (m^2_2 - \mu^2) \ln^2 (m^2_2/\mu^2)~.
\end{equation}

The total first-loop contribution to the on-shell $W$ -decay is therefore

 \begin{equation}\label{waddloop}
M^{(23)}_W = - M_W^{Born} \frac{\alpha ({Q'}_f^2 + \bar{Q}_f^2)}{8 \pi} L~
 \end{equation}
and accounting for the higher-loop DL contributions leads to the exponentiation of the first-loop contribution:

\begin{equation}\label{mw}
M_W = M_W^{Born} \exp \left[- \frac{\alpha ({Q'}_f^2 + \bar{Q}_f^2)}{8 \pi} L \right]~.
 \end{equation}
It is interesting to notice that when the $W$ -boson is highly off-shell, both graphs in Fig.~3 do not yield
DL contributions. Indeed, the propagator of the virtual $Z$ -boson  in Fig.~3a is
\begin{equation}\label{wprop}
(p_1 + p_2 - k)^2 - M^2 = \left[(p_1 + p_2)^2 - M^2 \right] + k^2 - 2p_1k - 2 p_2 k~.
\end{equation}
When $(p_1 + p_2)^2 \gg M^2 $, the first term in Eq.~(\ref{wprop}) is much greater than the other
terms and therefore they can be neglected.
It makes impossible to have DL contributions from graphs in Fig.~3 in this kinematic region.
As a result,  Eq.~(\ref{mw}) should be replaced in this kinematics by
\begin{equation}\label{woff}
M_W^{off} = M_W^{Born} \exp \left[\frac{\alpha {Q'}_f \bar{Q}_f}{4 \pi} L^{off} \right]~,
\end{equation}
with
\begin{equation}\label{loff}
L^{off} =  \ln^2 \left((p_1 + p_2)^2/\mu^2 \right) - \Theta (m^2_1 - \mu^2) \frac{1}{2} \ln^2(m^2_1/\mu^2)
- \Theta (m^2_2 - \mu^2) \frac{1}{2} \ln^2(m^2_2/\mu^2)~.
\end{equation}
Obviously, the exponential in Eq.~(\ref{woff}) steeply increases with the energy of the $W$ -boson. Had it been the
only contribution to the process, it would have led to violation of such fundamental principles as the unitarity.
As a matter of fact, this off-shell contribution is not gauge-invariant and therefore can be
regarded only as a sub-process of some more general reaction,
being supplemented by other DL contributions canceling the increase.

\section{Conclusion}
We have shown that there is no principal difference between the Sudakov form factors for on- and
off- shell neutral electroweak boson $Z$: in the both cases it is
\begin{equation}\label{zgen}
M_Z^{Born} \exp \left[- \frac{\alpha Q_f^2}{4 \pi} \ln^2 ((p_1 + p_2)^2/\mu^2) + \Theta (m^2_1 - \mu^2) \frac{1}{2} \ln^2(m^2_1/\mu^2)
+ \Theta (m^2_2 - \mu^2) \frac{1}{2} \ln^2(m^2_2/\mu^2) \right]
\end{equation}
while the behavior of the form factor of the charged $W$ -bosons is more involved:
it contains the negative exponent of Eq.~(\ref{mw}) when the $W$ is on-shell and
the positive exponent of Eq.~(\ref{woff}) when the $W$ is off-shell. This is caused by
the difference in behavior of graphs in Fig.~3: they yield DL terms only when the $W$ -boson is on-shell.
Considering the use of the Dispersion Relations, we have also shown that there is the big difference
in treatment of the infrared-divergent and regular contributions.

\section{Acknowledgement}

We are grateful to S.I.~Troyan and M.~Slawinska for useful discussions.

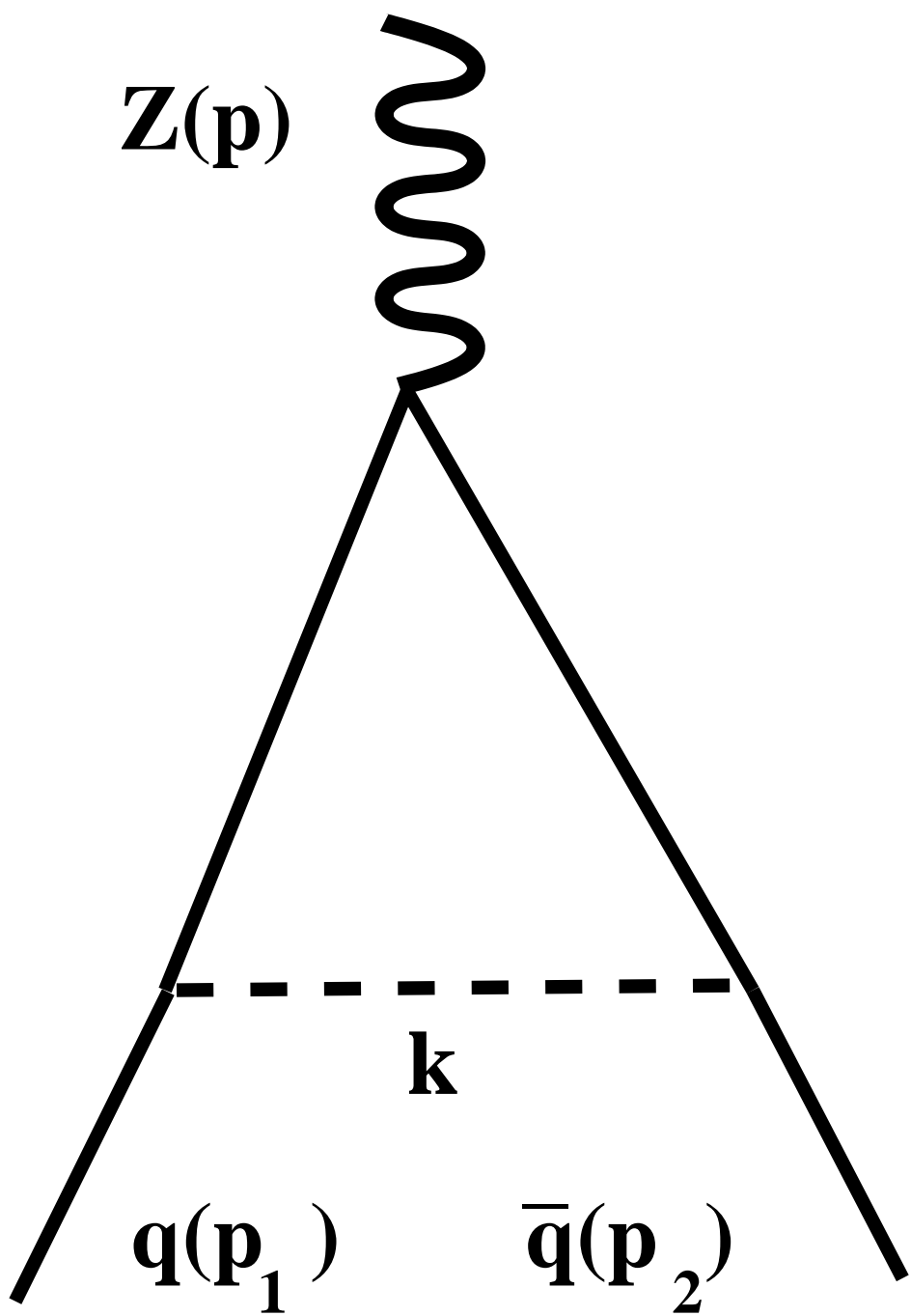
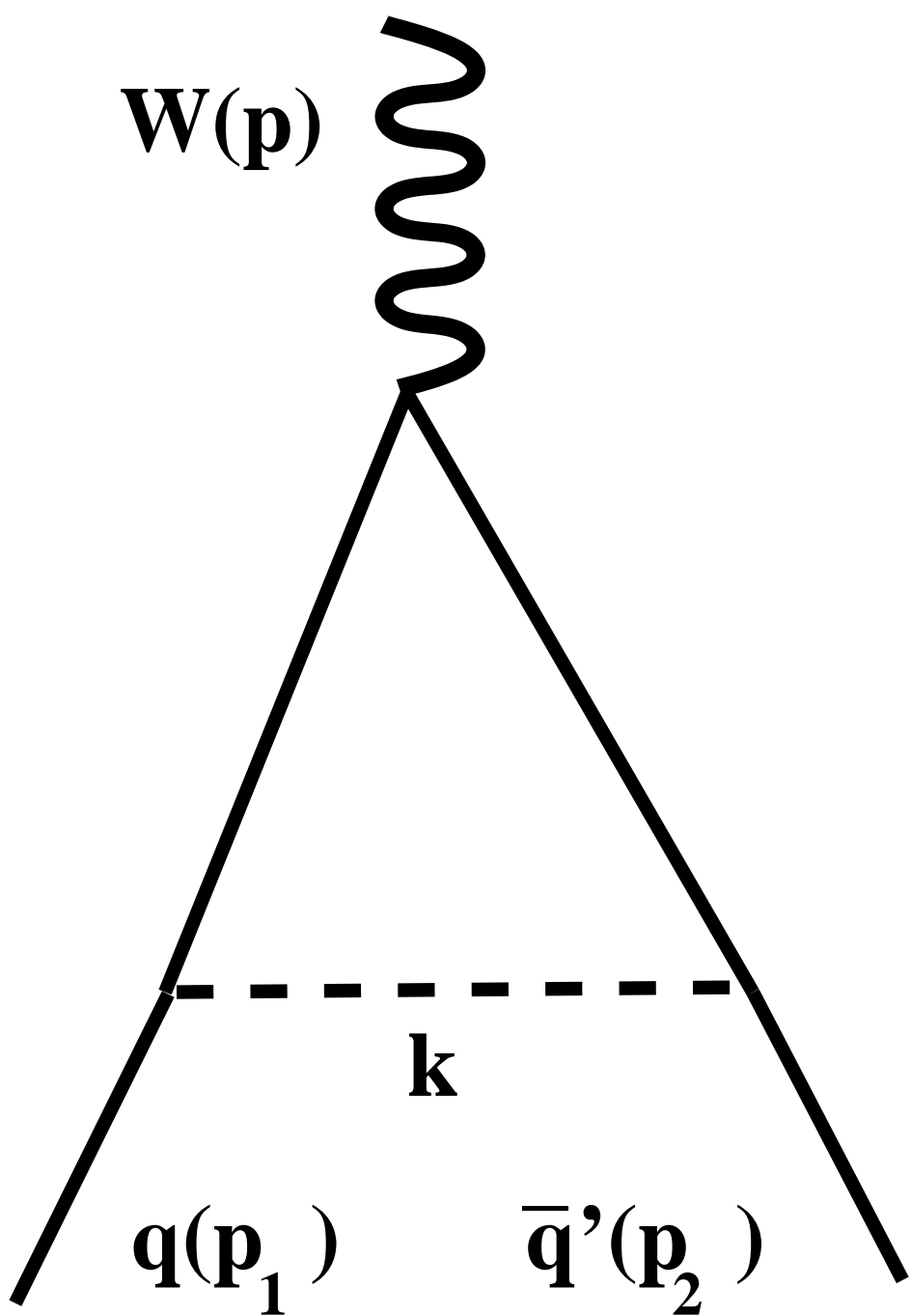
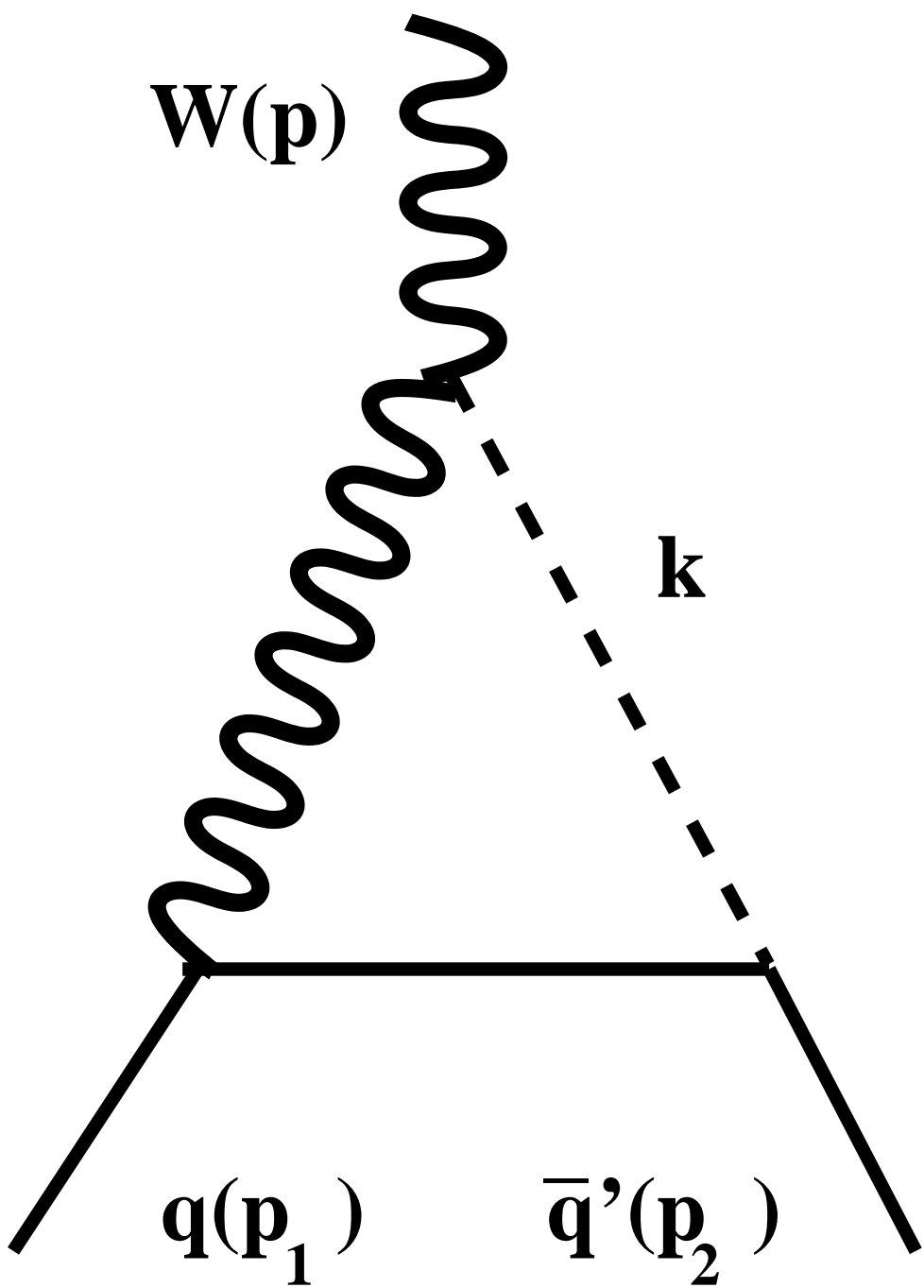
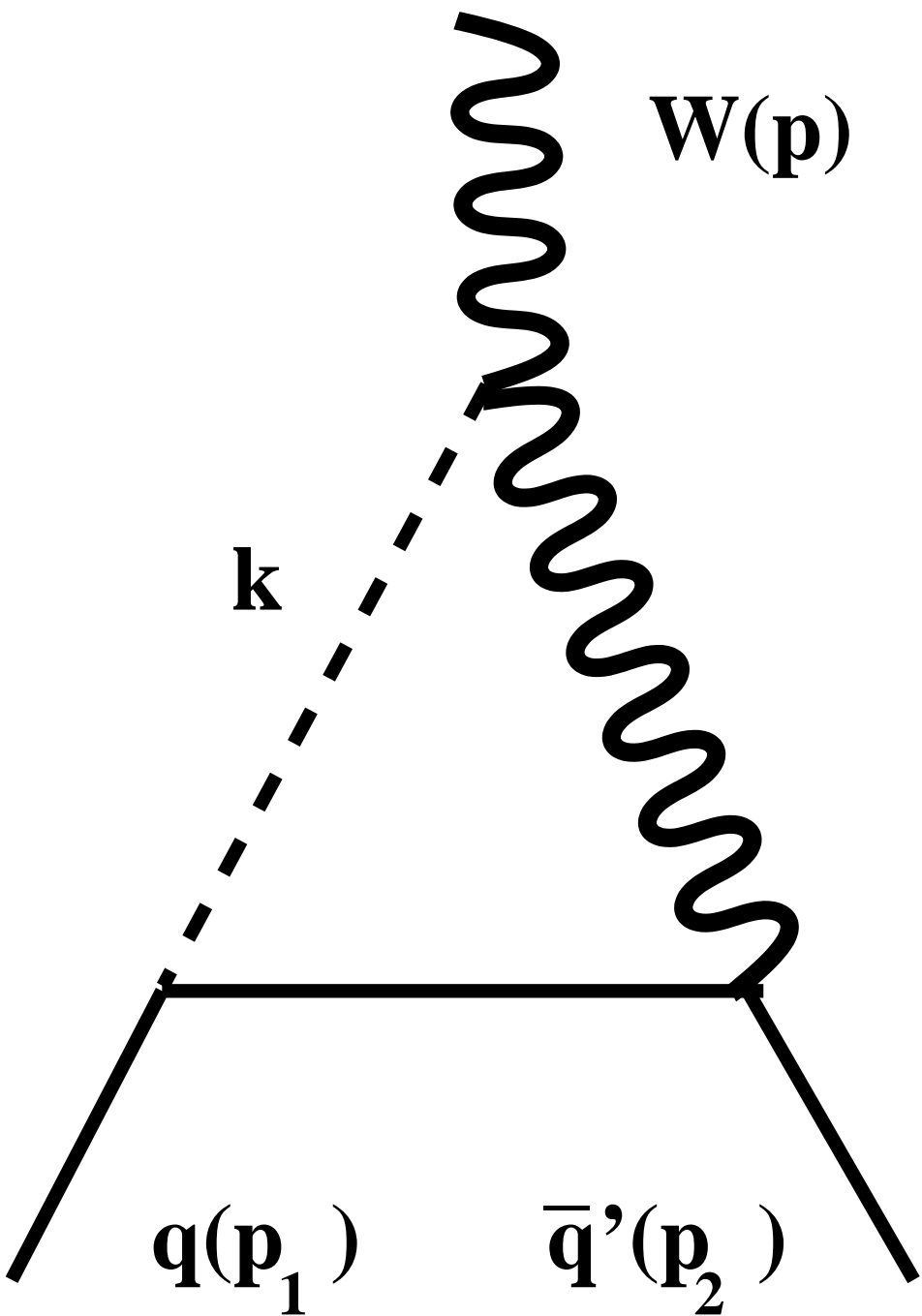
\end{document}